\documentclass[aps,twocolumn,prb,showpacs]{revtex4-1}
\usepackage{graphicx}
\usepackage{amsmath}
\usepackage{epstopdf}
\usepackage{amsbsy}
\usepackage{color}
\usepackage{lipsum}
\usepackage{dcolumn}
\usepackage{pstricks}
\usepackage{bm}

\def\frac#1#2{{\textstyle{#1 \over #2}}}

\def\be{\begin{equation}} \def\ee{\end{equation}}
\def\bea{\begin{eqnarray}} \def\eea{\end{eqnarray}}

\def\nn{\nonumber}

\def\vk{{\vec k}}

\def\vq{{\vec q}}
\def\vQ{{\vec Q}}

\begin{document}

\title{Structural and magnetic field effects on spin fluctuations in Sr$_{3}$Ru$_2$O$_7$}
\author{Shantanu Mukherjee}
\email{smukherj@binghamton.edu}
\affiliation{Department of Physics, Applied Physics, and Astronomy, Binghamton University - State University of New York, Binghamton, USA}

\author{Wei-Cheng Lee}
\email{wlee@binghamton.edu}
\affiliation{Department of Physics, Applied Physics, and Astronomy, Binghamton University - State University of New York, Binghamton, USA}

\date{\today}

\begin{abstract}
We investigate the evolution of magnetic excitations in Sr$_3$Ru$_2$O$_7$ using a three band tight binding model that takes into account the influence of Mn and Ti dopant ions. The effect of dopant ions on the Sr$_3$Ru$_2$O$_7$ band structure has been included by taking into account the dopant induced suppression of the oxygen octahedral rotation in the tight binding band structure and changes in electron occupation. We find that the low energy spin fluctuations are dominated by three wave vectors around $\vQ=((0,0),(\pi/2,\pi/2))$, and $(\pi,0)$ which compete with each other. As the octahedral rotation is suppressed with increasing doping, the three wave vectors evolve differently. In particular, the undoped compound has dominant wavevectors at $\vQ=((0,0),(\pi/2,\pi/2))$, but doping Sr$_3$Ru$_2$O$_7$ leads to a significant enhancement in the spin susceptibility at the $\vQ=(\pi,0)$ wavevector bringing the system closer to a magnetic instability. All the features calculated from our model are in agreement with neutron scattering experiments. We have also studied the effect of a c-axis Zeeman field on the low energy spin fluctuations. We find that an increasing magnetic field suppresses the AFM fluctuations and leads to stronger competition between the AFM and FM spin fluctuations. The magnetic field dependence observed in our calculations therefore supports the scenario that the observed nematic phase in the metamagnetic region in Sr$_3$Ru$_2$O$_7$ is intimately related to the presence of a competing ferromagnetic instability.
\end{abstract}


\maketitle

{\it Introduction} --
Bilayer compound Sr$_3$Ru$_2$O$_7$ (Sr327) is a member of the Ruddlesden Popper series.\cite{ruddlesden:1958} Although it is not superconducting like single layer Sr$_2$RuO$_4$,\cite{maeno:1994} the phase diagram of  Sr$_{3}$Ru$_2$O$_7$ is quite rich. The undoped compound is a meta-magnetic metal on the verge of ferromagnetism and shows a region of significant change in resistivity \cite{perry:2001,grigera:2001,ikeda:2000}  and development of a spin density wave order in moderate magnetic fields ($B\sim8$ T magnetic field applied along the c-axis).\cite{lester:2015} Rotating the field direction away from the c-axis affects the meta-magnetic transition and a nematic state has been observed that is characterized by a strong anisotropy in resistivity within the ab-plane.\cite{lester:2015,bruin:2013} Apart from the unusual magnetic field behavior, doping  Sr$_{3}$Ru$_2$O$_7$ with Manganese (Mn) or Titanium (Ti) also leads to a number of interesting properties. Both Mn and Ti act as substitutional impurities and replace the Ru ions. Above a critical doping concentration of $x\sim5\%$,  lowering the temperature leads to an insulating state followed by a long range anti-ferromagnetic order.\cite{mathieu:2005,hossain:2013} The metal insulator transition temperature grows with doping whereas the long range AFM state shows a dome like behavior in the temperature-doping phase diagram and exists upto a doping level of $x\sim20\%$.\cite{mathieu:2005,hossain:2008,steffens:2009} Small amounts of Ti dopants in Sr327 suppress the meta-magnetic transition and this suppression has been argued to result from competing magnetic interactions in Sr327.\cite{hooper:2007} Therefore understanding the low energy spin fluctuations in Mn or Ti doped Sr327 compounds will be crucial for unraveling the origin of dopant induced magnetic order as well as the physics of meta-magnetic transition. 

Undoped Sr327 is non-magnetic but shows low energy spin fluctuations at $\vQ_{AFM}=(Q_1,Q_2)=(\pi/2,\pi/2)$  wavevector.\cite{capogna:2003} Note that we will be working in the larger orthorhombic unit cell.\footnote{To convert to tetragonal cell use $Q_x=(Q_1-Q_2)/2$, and $Q_y=(Q_1+Q_2)/2$.} Inelastic neutron scattering experiments also observe enhanced incommensurate ferromagnetic spin fluctuations at $\vQ_{FM}\sim(0.09,0.09,0)$.\cite{capogna:2003} These features in the paramagnetic undoped compound can be understood from general Fermi surface nesting arguments\cite{singh:2001}. However upon doping with Mn, neutron scattering experiments find a long range AFM state with dominant wave vector at $\vQ \sim (\pi,0)$\cite{mesa:2012} and doping with iso-valent Ti$^{4+}$ ions leads to an AFM state with incommensurate wavevector $\vQ \sim (\pi+\delta,0)$\cite{steffens:2009}. Therefore, this strong shift in the ordering vector of low energy fluctuations cannot be explained simply by a rigid band shift induced by Mn doping. 

Structurally, Sr$_3$Ru$_2$O$_7$ compound differs from Sr$_2$RuO$_4$ not only by the presence of a bilayer coupling but also due to a significant oxygen octahedral rotation of $\theta\sim 7^{\circ}$.\cite{huang:1998} The band folding associated with the rotated octahedra leads to a complex Fermi surface and band structure in Sr$_3$Ru$_2$O$_7$\cite{tamai:2008}. The structural parameter controlling the degree of octahedral rotation is strongly affected by the presence of Mn dopants. It has been observed in x-ray diffraction experiments that upon increasing the Mn concentration, the octahedral rotation is strongly suppressed by a doping of $x\sim 16\%$.\cite{hu:2011,li:2013} Therefore in order to identify the mechanism governing the magnetic order we need to study the effects of changes in octahedral rotation, and electron occupation on the low energy spin fluctuations.

In this work, we use a three orbital tight binding model with spin orbit interaction and octahedral rotation applicable to Sr327 and calculate the dynamical spin susceptibility $\chi(\vQ,\omega)$ by including electron correlations within the random phase approximation (RPA). This model has been used to obtain the nematic order in undoped Sr327 compounds.\cite{puetter:2010,lee2:2010} We model the doping dependence on this system by the suppression of a single tight binding parameter that simulates the effect of Mn doping on the octahedral rotation. We find that the imaginary part of spin susceptibility indeed shifts from  $\vQ \sim (\pi/2,\pi/2)$ in the undoped compound to $\vQ \sim (\pi,0)$ near optimal doping. Additionally, we find the spin fluctuations are also strongly enhanced in magnitude as octahedral rotation gets suppressed indicating a tendency towards a magnetic order at the $\vQ \sim (\pi,0)$ wave vector observed in experiments. This provides evidence that the magnetic transition in Mn doped Sr$_3$Ru$_2$O$_7$ is primarily governed by the doping induced changes in the lattice structure. We further calculate the effect of a c-axis Zeeman field on the spin fluctuations in the undoped compound and find that the AFM fluctuations not only become more incommensurate but also get progressively weaker near the $\vQ \sim (\pi/2,\pi/2)$ region. This magnetic field behavior would lead to a stronger competition between the AFM and FM fluctuations in the undoped Sr327 compound. 
\begin{figure}	
	\rput[Bl](-0.05\columnwidth,0.01\columnwidth){(a)}
	\includegraphics[width=0.45\columnwidth]{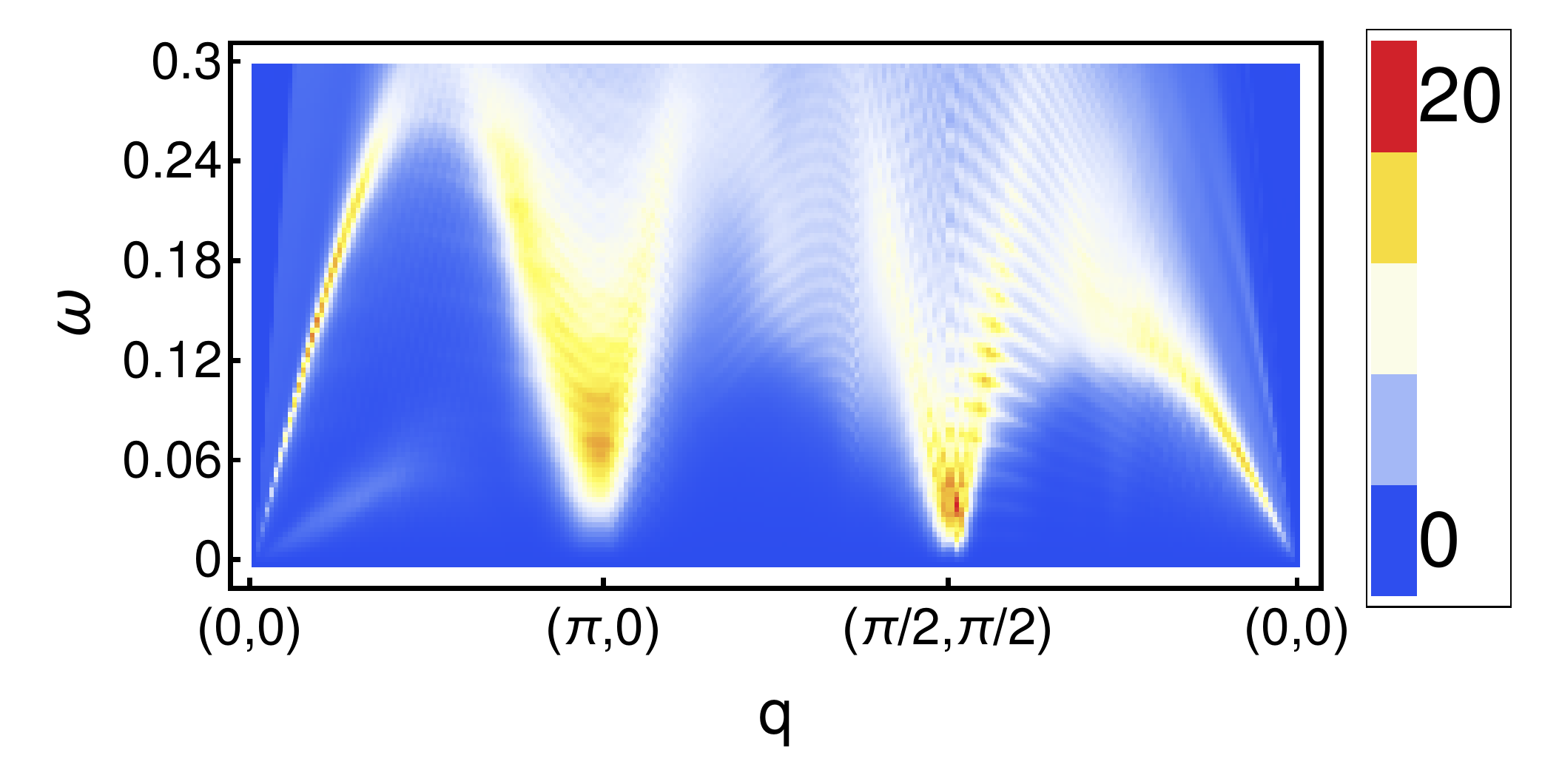} 
	\rput[Bl](0.02\columnwidth,0.01\columnwidth){(b)}
	\includegraphics[width=0.45\columnwidth]{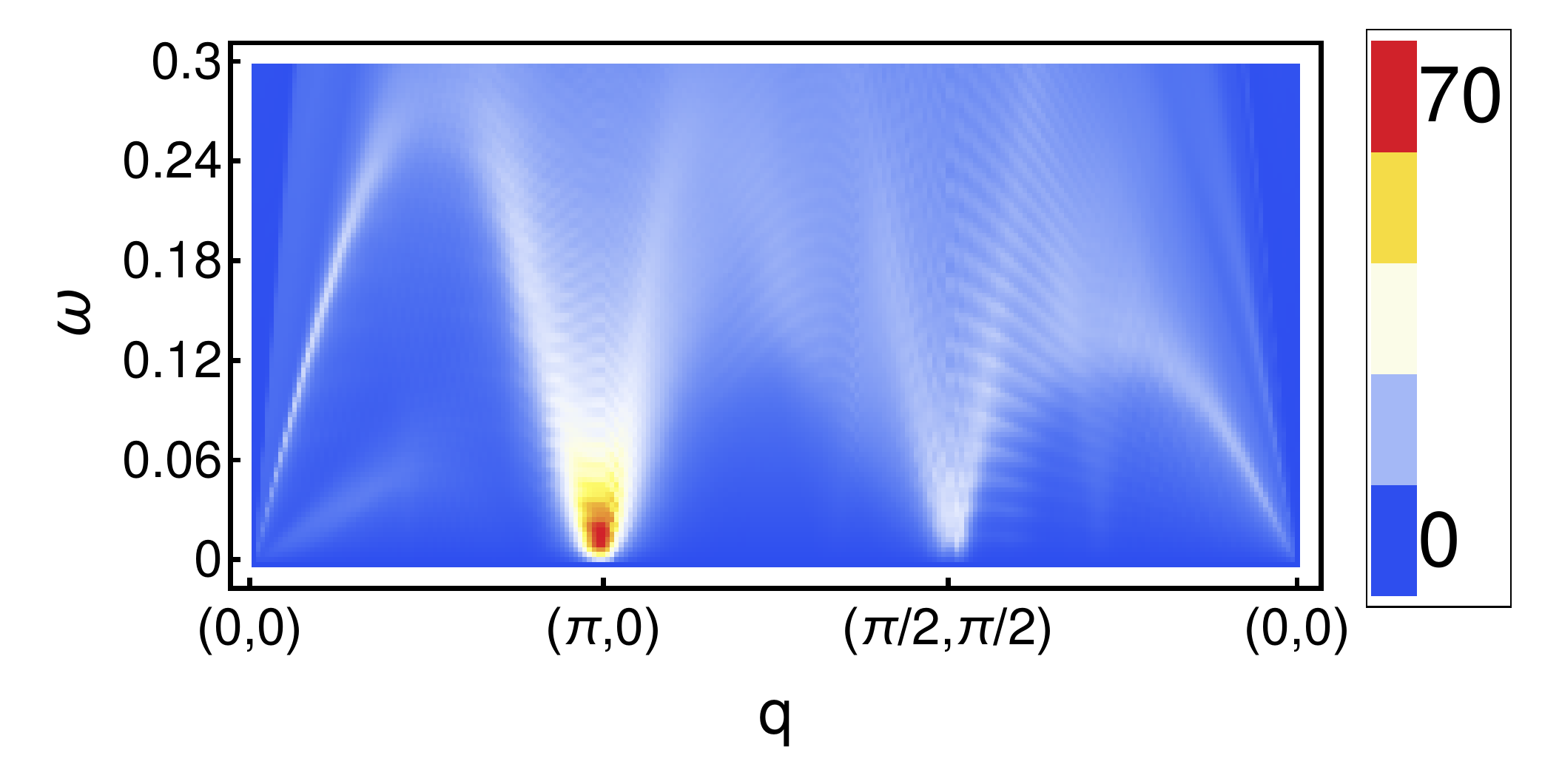}
	\rput[Bl](-0.05\columnwidth,0.01\columnwidth){(c)}
	\includegraphics[width=0.45\columnwidth]{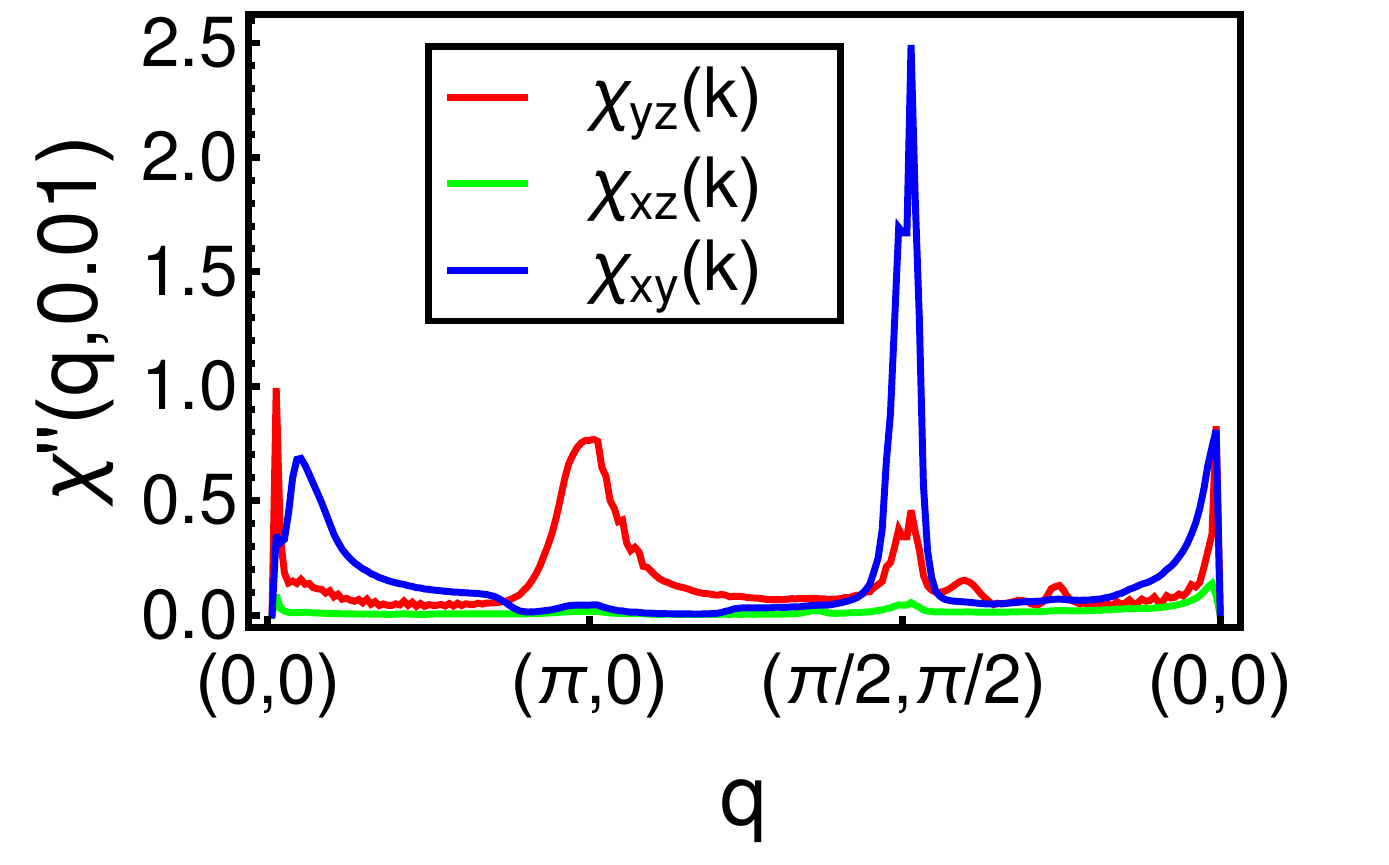}
	\rput[Bl](0.02\columnwidth,0.01\columnwidth){(d)}
	\includegraphics[width=0.45\columnwidth]{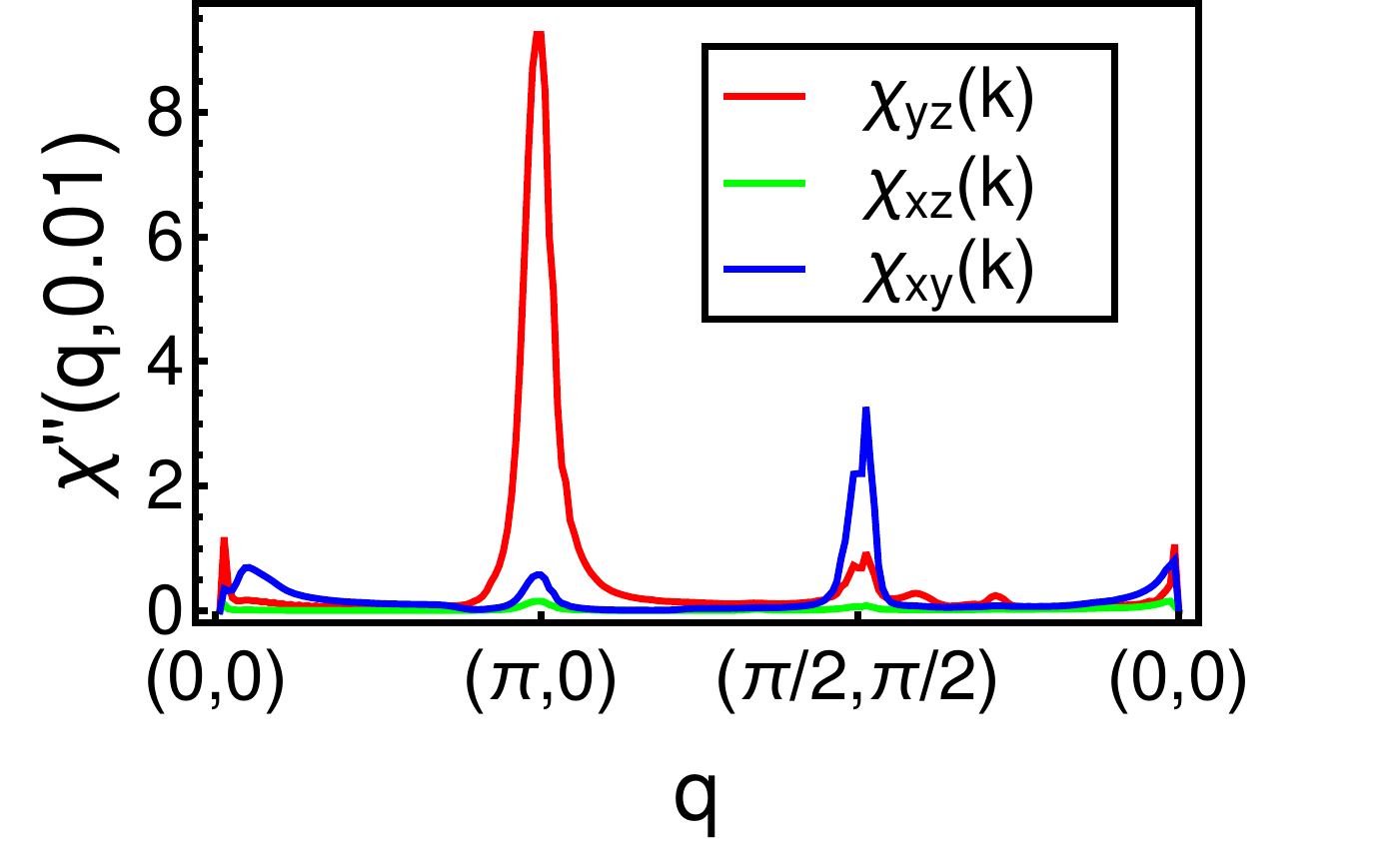}

	\caption{\label{fig:imchi} Plots showing the imaginary part of RPA spin susceptibility $\chi''_{RPA}(\vq,\omega)$ for $x=1\%$ (a) and $x=6\%$ (b) Mn doping. The corresponding orbital resolved spin fluctuation components $\chi_{aaaa}^{zz}(\vQ,0.01)$ where $a=(d_{yz},d_{xz},d_{xy})$ orbitals are shown for $x=1\%$(c) and $x=6\%$(d) doping.  They have been calculated for $U=0.8$, and $J=0.25U$.}
\end{figure}

{\it Model} --
We model the electronic structure of Sr$_3$Ru$_2$O$_7$ by using a 3 orbital tight binding model consisting of $t_{2g}$ orbitals ($d_{xz},d_{yz},d_{xy}$) which are most relevant for its low energy properties. The tight binding Hamiltonian considered by us has been discussed in Ref.~\onlinecite{puetter:2010} and effects of octahedral rotation as well as interlayer coupling are given in Ref.~\onlinecite{lee:2010}. This model captures the essential features of the band structure of Sr327 including the Fermi surface (see Fig S1 in supplementary section), density of states, and the Van Hove singularity that is believed to play an important role in the low energy properties of this compound. The Hamiltonian can be written as,
\be
H=H_{0}+H_{INT}
\ee
The tight binding Hamiltonian $H_{0}$ for the bilayer model can be written in terms of two independant parts after Fourier transforming along the layer index\cite{lee:2010}. Writing the bilayer Hamiltonian as $H_{0}=h_0(k_z=0)+h_0(k_z=\pi)$, each component is given by,
\be
h_{0}(k_z)=\sum_{\vk}\Phi^{\dag}_{\vk,s,k_z}
\left(\begin{array}{ll}
\hat{h}_{0s}(\vk,k_z) & \hat{g}^{\dag}(\vk,k_z)\\
\hat{g}(\vk,k_z) & \hat{h}_{0s}(\vk+\vQ,k_z)
\end{array}\right)
\Phi_{\vk,s,k_z}
\label{eq:h0}
\ee
where,
\be 
\Phi_{\vk,s,k_z}=(d^{yz}_{\vk,s,k_z},d^{xz}_{\vk,s,k_z},d^{xy}_{\vk,-s,k_z},
d^{yz}_{\vk',s,k_z},d^{xz}_{\vk',s,k_z},d^{xy}_{\vk',-s,k_z})
\ee 
 with $\vk'=\vk+\vQ$, and $d^{\alpha}_{\vk,s,k_z}$ annihilates an electron with orbital $\alpha$, spin $s$, in-plane momenta $\vk$, and momentum along $z$-direction $k_z$. The matrix components are given by,
\bea
\hat{h}_{0s}(\vk,k_z)=\hat{A}_s(\vk)+\hat{B}_1\cos(k_z)\\
\hat{g}(\vk,k_z)=\hat{G}(\vk)-2\hat{B}_2\cos(k_z)
\eea

The $3\times 3$ matrix kernel are the intra-layer hopping contribution without octahedral rotation $\hat{A}_s(\vk)$, staggered in-plane hopping $\hat{G}(\vk)$, inter-layer hopping without rotation $\hat{B}_1$, and staggered inter-layer hopping $\hat{B}_2$ which have been discussed in Ref.~\onlinecite{lee:2010}. The matrix elements are given by,

\be
\hat{A_s}(\vk)=
\left(\begin{array}{lll}
	\epsilon_{\vk}^{yz} & \epsilon_{\vk}^{off}+is\lambda & 0\\
\epsilon_{\vk}^{off}-is\lambda & \epsilon_{\vk}^{xz} & 0\\
	0 & 0 & \epsilon_{\vk}^{xy}
\end{array}\right)
\ee
The individual matrix elements for $\hat{A}_s(\vk)$, $\hat{G}(\vk)$, $\hat{B}_1$, and $\hat{B}_2$ are provided in the Supplementary material. Note that the spin orbit coupling $\lambda$ is assumed to primarily couple the $d_{xz/yz}$ orbitals and ignores the coupling to the $d_{xy}$. A similar spin orbit coupling term has been argued to explain the low energy fluctuations in Sr$_2$RuO$_4$.\cite{eremin:2002} 

We use the octahedral rotation hopping matrix element $t_{rot}$ as a free parameter to simulate the effect of Mn doping on the Sr327 compound. Although we also include a weak bilayer splitting of the bands by including $\hat{B}_1$ and $\hat{B}_2$ matrices, they are kept fixed as a function of doping. This is reasonable since they do not significantly affect the doping dependence of the low energy spin fluctuations.

Finally the Zeeman term includes the effect of an external magnetic field B oriented along the out of plane c-axis direction. It is given by,
\be
H_{Zeeman}= B\sum_{s,\vk,\alpha,a}(-1)^{s}d^{\dag \alpha}_{s,a}(\vk)d^{\alpha}_{s,a}(\vk)
\ee

Here the energy scale for the Zeeman term $B$ is expressed in units of hopping $t$. The electron correlations are included in our model using an on site Hubbard Hund interaction. Following the discussion in Ref~\onlinecite{wu:2015} the interaction Hamiltonian in real space can be written as,
\bea
H_{INT}=\sum_{i\mu\nu\theta\psi s_1s_2}U_{\theta\psi s_2}^{\mu\nu s_1}d_{i\mu s_1}^{\dag}d_{i\nu s_1}d_{i\theta s_2}^{\dag}d_{i\psi s_2}
\eea
where $i$ is the site index, ($\mu,\nu,\theta,\psi$) are orbital indices and $s_1,s_2$ are the spin indices. The components of the interaction matrix are given in Ref.~\onlinecite{wu:2015}. 

The longitudinal spin susceptibility within RPA is a $36\times36$ matrix given by,
\bea
&&\hat{\chi}^{zz}_{RPA}(\vq,i\omega_n)=[\hat{1}+\hat{M}_{22}^{-1}\hat{M}_{21}]\nn\\
&&[\hat{1}-(\hat{M}_{11}^{-1}\hat{M}_{12})((\hat{M}_{22}^{-1}\hat{M}_{21})]^{-1}\hat{M}_{11}^{-1}\hat{\chi}^{\uparrow\uparrow}(\vq,i\omega_n)+\nn\\
&&[\hat{1}+\hat{M}_{11}^{-1}\hat{M}_{12}][\hat{1}-(\hat{M}_{22}^{-1}\hat{M}_{21})(\hat{M}_{11}^{-1}\hat{M}_{12})]^{-1}\nn\\
&&\hat{M}_{22}^{-1}\hat{\chi}^{\downarrow\downarrow}(\vq,i\omega_n)\\
&& \hat{M}=
\left(\begin{array}{ll}
	\hat{1}+4\hat{\chi}^{\uparrow\uparrow}(\vq,i\omega_n)\hat{U} & 2\hat{\chi}^{\uparrow\uparrow}(\vq,i\omega_n)\hat{V}\\
	2\hat{\chi}^{\downarrow\downarrow}(\vq,i\omega_n)\hat{V} & 	\hat{1}+4\hat{\chi}^{\downarrow\downarrow}(\vq,i\omega_n)\hat{U}
\end{array}\right)
\eea
and the transverse susceptibility within RPA can be written as,
\be
\hat{\chi}^{\pm}_{RPA}(\vq,i\omega_n)=[\hat{1}+\chi^{\uparrow\downarrow}(\vq,i\omega_n)(4\hat{U}-2\hat{V})]^{-1}\chi^{\uparrow\downarrow}(\vq,i\omega_n)
\ee
Where the bare susceptibility for the spin up channel can be expressed as,
\bea
&&\chi_{l_1l_2l_3l_4}^{\uparrow\uparrow}(\vq,i\omega_n)=\nn\\
&&{1\over N}\sum_{\vk\alpha\beta}W_{l_4l_1l_2l_3}^{\alpha\beta}(\vk,\vq){n_F(\epsilon_{\vk+\vq}^{\beta})-n_F(\epsilon_{\vk}^{\alpha})\over i\omega_n+(\epsilon_{\vk}^{\alpha}-\xi_{\vk+\vq}^{\beta})}\\
&&W_{l_4l_1l_2l_3}^{\alpha\beta}(\vk,\vq)\nn\\
&&=\psi_{l_4\uparrow}^{\alpha}(\vk)\psi_{l_1\uparrow}^{\alpha*}(\vk)\psi_{l_2\uparrow}^{\beta}(\vk+\vq)\psi_{l_3\uparrow}^{\beta*}(\vk+\vq)
\eea

In the absence of a Zeeman field the other components of the bare susceptibility can be expressed as,
\bea
&&\chi_{l_1l_2l_3l_4}^{\downarrow\downarrow}(\vq,i\omega_n)=\chi_{l_4l_3l_2l_1}^{\uparrow\uparrow}(\vq,i\omega_n)\\
&&\chi_{l_1l_2l_3l_4}^{\uparrow\downarrow}(\vq,i\omega_n)=\chi_{l_4l_2l_3l_1}^{\uparrow\uparrow}(\vq,i\omega_n)
\eea

{\it Results} --
The most notable structural effect of Mn doping is the suppression of octahedral rotation.\cite{hu:2011} We have modeled the effect of Mn doping in Sr$_3$(Ru$_{1-x}$Mn$_x$)$_2$O$_7$ by varying the hopping parameter $t_{rot}$ that represents the influence of octahedral rotation on the tight binding Hamiltonian. For $t_{rot}\sim 0.4$ ($t_{rot}\sim 0.2$) leads to an electron doping of $n\sim 4.01e^{-}/$Ru ($4.15e^{-}/$Ru) with $x\propto n$ for an Mn dopant ion. We would later discuss the effect of an isovalent dopant such as Ti on the spin fluctuations.

\begin{figure}
	\rput[Bl](0.2\columnwidth,0.03\columnwidth){(a)}
	\includegraphics[width=0.49\columnwidth]{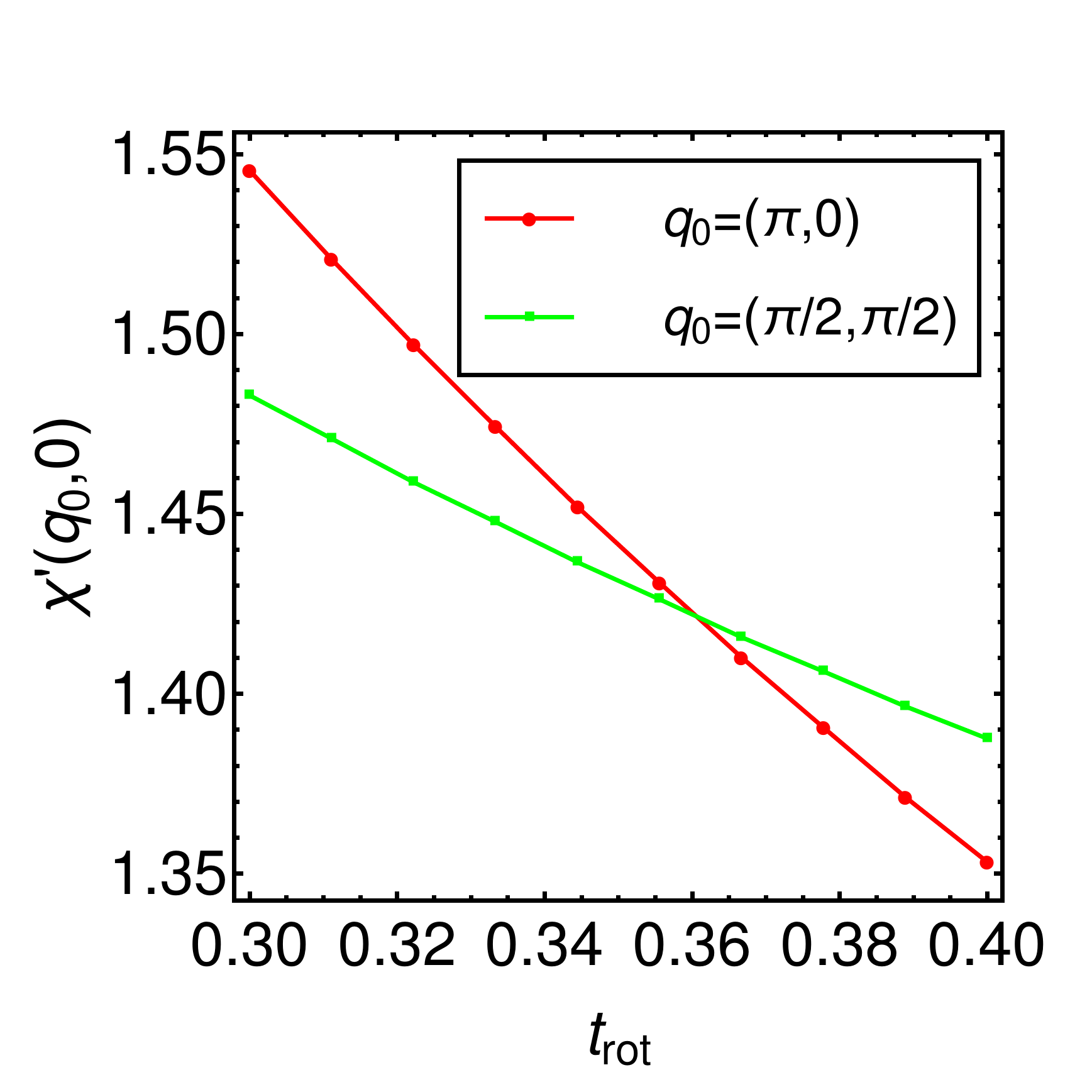}
	\rput[Bl](0.2\columnwidth,0.03\columnwidth){(b)}
	\includegraphics[width=0.49\columnwidth]{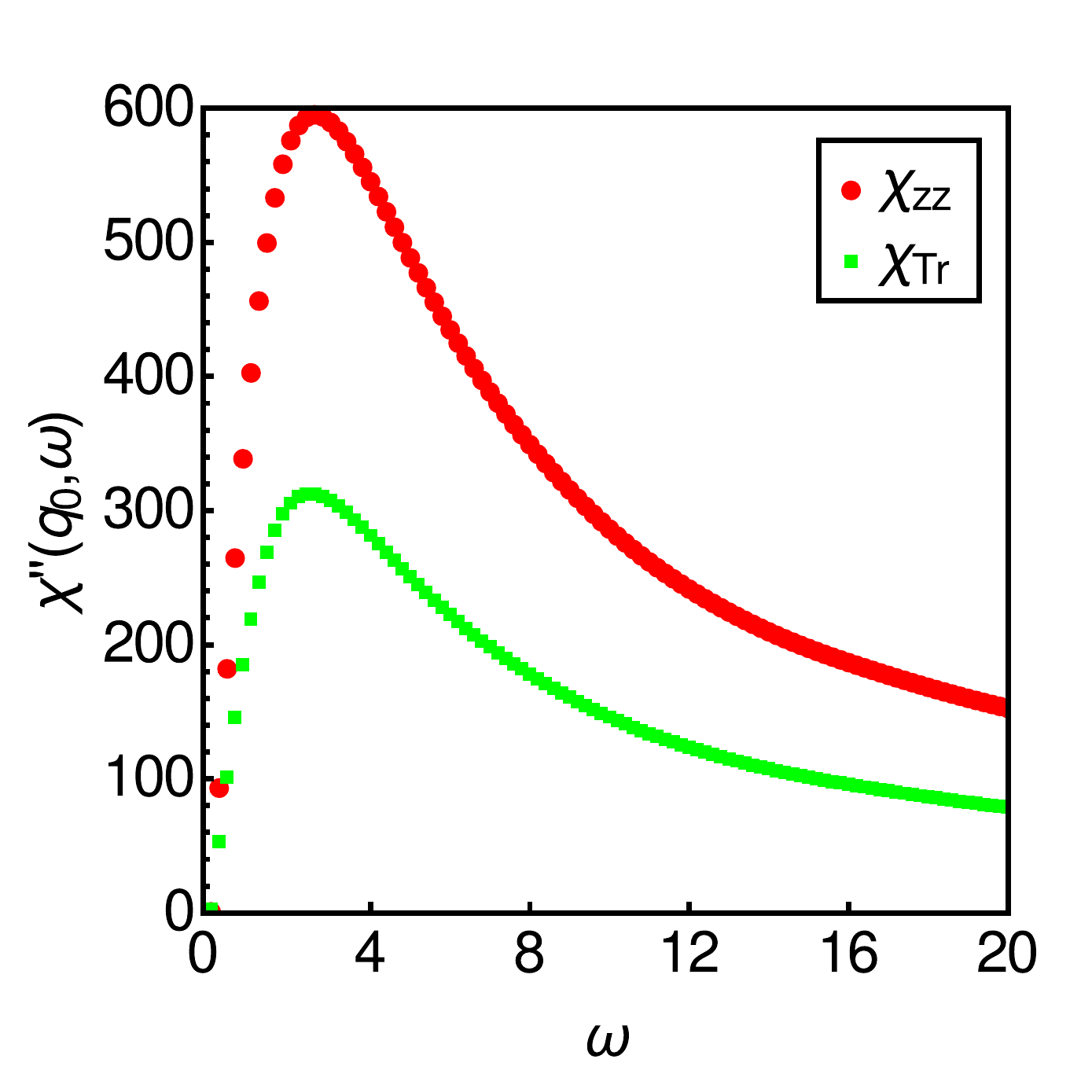}
	\caption{\label{fig:dopdep} (a) The Real part of spin susceptibility at $\vQ=(\pi,0)$ and $(\pi/2,\pi/2)$ as a function of increasing $t_{rot}$ (or decrease in doping $x$). (b) The imaginary part for longitudinal and transverse susceptibility at $x=1\%$ doping and $\vQ=(\pi/2,\pi/2)$ as a function of energy. }
\end{figure}

The calculated dynamical susceptibility in the longitudinal channel $\chi^{''}_{RPA}(\vQ,\omega)$ is presented in Fig.~\ref{fig:imchi} for different doping levels. In Fig.~\ref{fig:imchi}a we show $\chi''_{RPA}(\vQ,\omega)$ for $t_{rot}=0.4$ and corresponding to an Mn doping of $x\sim 1\%$. The dominant low energy spin fluctuations are peaked at $\vQ=(\pi/2,\pi/2)$ wavevector in agreement with neutron scattering experiments\cite{capogna:2003}. The spin fluctuations in the undoped compound has a large contribution at $\vQ=(\pi/2,\pi/2)$ from the intra orbital susceptibility corresponding to the $d_{xy}$ band (see Fig.~\ref{fig:imchi}(c)). The $(\pi/2,\pi/2)$ nesting contribution from the $d_{xy}$ band is enhanced because of the presence of a Van-Hove singularity in the $d_{xy}$ band of the undoped compounds. Note that for the $x\sim1\%$ compound in Fig.~\ref{fig:imchi}, there is a significant sub-dominant contribution to the susceptibility at $\vQ=(\pi,0)$ which can also be seen in the static spin susceptibility $\chi'(\vQ,0)$ shown in  Fig.~\ref{fig:dopdep}(a). However being weaker than the susceptibility at $\vQ=(\pi/2,\pi/2)$, it is relatively suppressed for larger Coulomb interaction closer to the Stoner instability. In addition to the AFM fluctuations we also find strong ferro-magnetic spin fluctuations at low energies which disperse at higher energies. Interestingly the relative strength of the AFM and FM fluctuations depends on the strength of the Coulomb interactions. We find that at low energies the FM fluctuations dominate in the imaginary part of the bare susceptibility, but are weaker than AFM fluctuations when electron correlations are included in the model.   

In Fig.~\ref{fig:imchi}b we show the calculated $\chi''(\vQ,\omega)$ for a weaker octahedral rotation $t_{rot}=0.36$ which corresponds to a doping of $x\sim 6\%$. The suppression of octahedral rotation not only shifts the spin fluctuations from $\vQ=(\pi/2,\pi/2)$ towards $\vQ=(\pi,0)$ but also enhances them significantly. This is evident from the doping dependence of $\chi'(\vQ,0)$ plotted in Fig.~\ref{fig:dopdep}(a) which shows that as the octahedral rotation gets reduced there is a crossover from $\vQ=(\pi/2,\pi/2)$ to $\vQ=(\pi,0)$ in the susceptibility at $t_{rot}\sim0.355$ or $x\sim5\%$. As can be seen from Fig.~\ref{fig:imchi}(d), the peak at $\vQ=(\pi,0)$ in doped Sr327 is primarily due to the $d_{xz/yz}$ intra-orbital nesting channel unlike the undoped compound where the dominant fluctuations correspond to the $d_{xy}$ band. Note that if the electron doping is kept fixed at $n=4.01e^{-}/$Ru, then at $t_{rot}=0.36$ we find that the spin fluctuations are still enhanced but become incommensurate (see supplementary Fig.~S2). This observation may be relevant to observations of incommensurate wave vector seen in neutron experiments on Ti doped Sr327 since the Ti dopants have the same oxydation state as Sr. \cite{steffens:2009} 

The enhancement in the susceptibility would bring the system close to a stoner instability as the octahedral rotation is progressively reduced. Note that under this scenario local magnetic order can already be expected to form near Mn dopants before the overall system crosses the stoner instability. Such behavior has been observed in REXS experiments\cite{hossain:2013}. We therefore argue that the $\vQ=(\pi,0)$  magnetic state observed in Mn doped Sr327 compounds is caused by a doping induced structural distortion that leads to a change in nesting properties at the Fermi surface. These dominant spin fluctuations are already present in the high temperature metallic phase. 

In Fig.~\ref{fig:dopdep}(b) we show the energy dependence of the dynamical susceptibility for the undoped system along the longitudinal and transverse direction. This anisotropy in the spin channel is caused by a spin orbit coupling and leads to an enhancement in the longitudinal bare susceptibility by a factor of 2, which has been argued to be the result of larger number of scattering channels along the longitudinal direction.\cite{eremin:2002} The energy dependence shows a peak at low energies similar to observations in neutron scattering experiments.\cite{capogna:2003} Such a peak at low energies in itinerant models usually indicates the proximity of the system to a stoner instability and would not exist if the system is far from a magnetic instability. 

\begin{figure}
	\rput[tr](-0.3\columnwidth,0.02\columnwidth){(a)}
	\includegraphics[width=0.49\columnwidth]{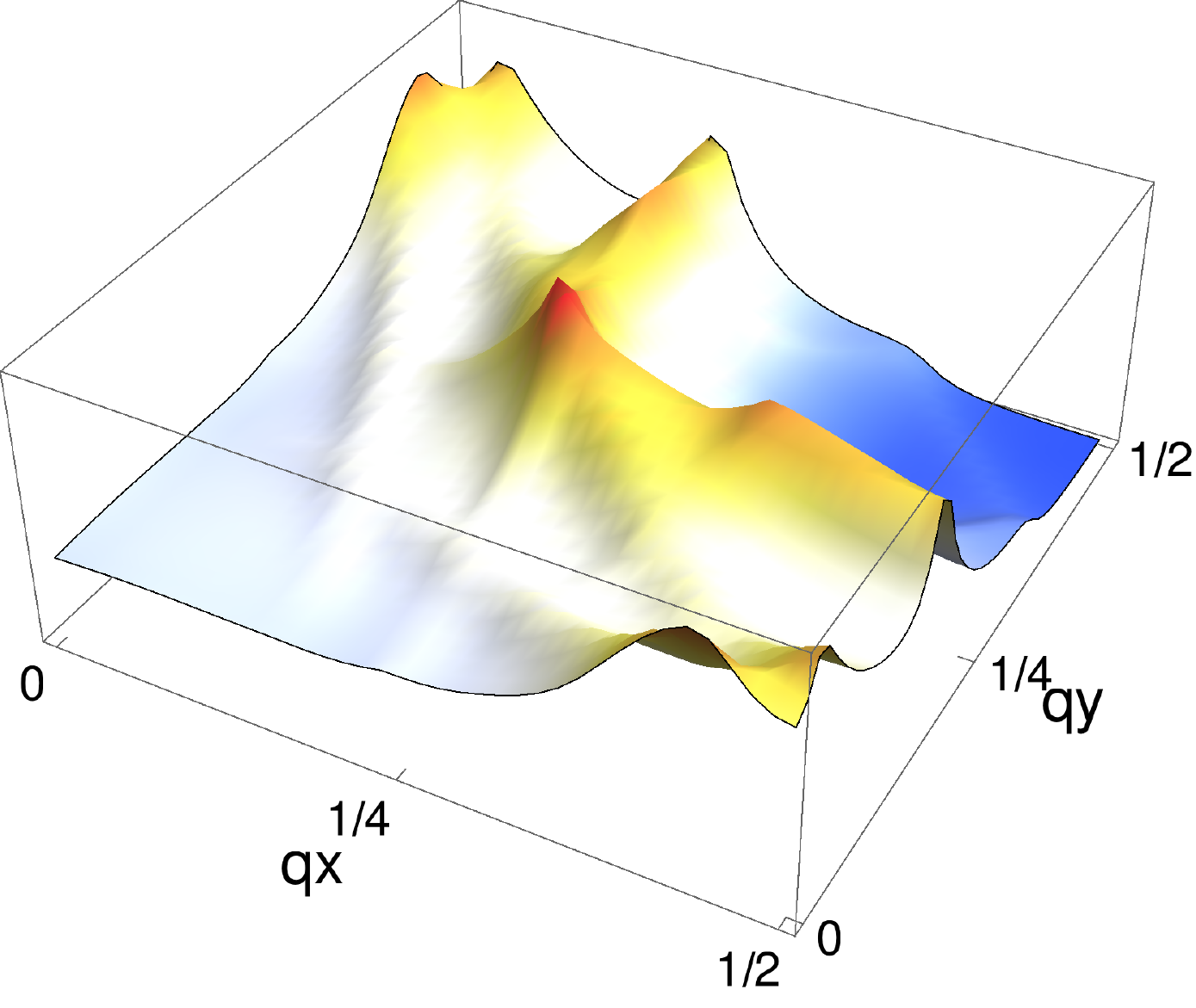}
	\rput[tr](-0.3\columnwidth,0.02\columnwidth){(b)}
	\includegraphics[width=0.49\columnwidth]{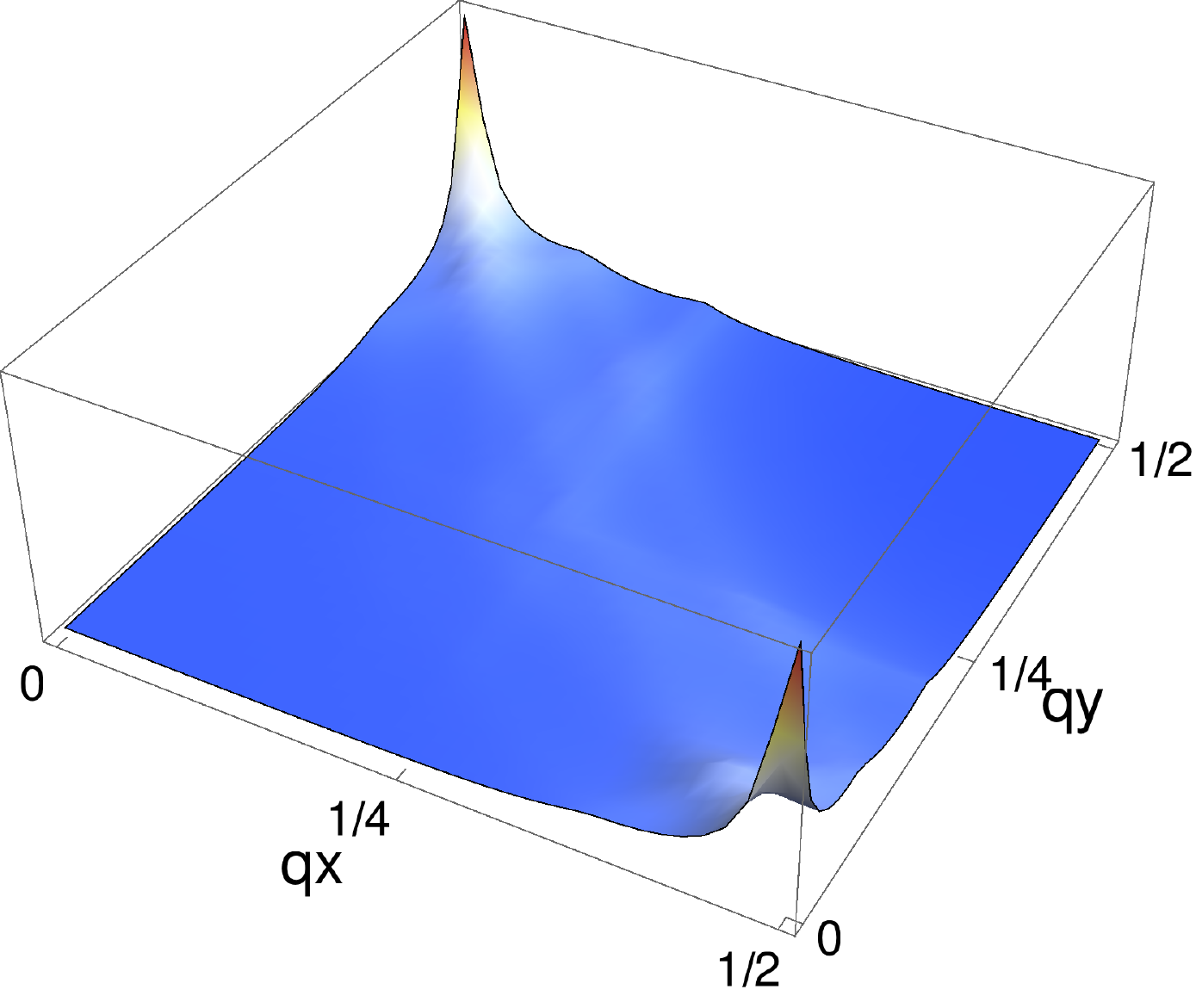}
	\caption{\label{fig:rechi} The real part of susceptibility $\chi^{'}_{RPA}(\vQ,0)$ for U=0.8, J=0.25U. The plots correspond to (a) $x\sim1\%$ and (b) $x\sim6\%$.The coordinates are given in the orthorhombic unit cell in units of $2\pi/a$.} 
\end{figure}

The real part of magnetic susceptibility $\chi'(\vQ,0)$ is shown in Fig.~\ref{fig:rechi}a (Fig.~\ref{fig:rechi}b) for $x\sim1\%$ ($x\sim6\%$). Note that for the lower doping, in addition to the $\vQ=(\pi/2,\pi/2)$ peak in the susceptibility, there is significant weight around the $(\pi,0)$ regions and along the $(\pi/2,\pi/2)-(\pi/2,0)$ line. These pseudo one dimensional (1D) nesting regions originate from the squarish $\Gamma$-centered Fermi pockets (see Fig.~~S1). For $x\sim6\%$, in Fig.~\ref{fig:rechi}b the static susceptibility is much stronger and dominated by the $(\pi,0)$ nesting. Note that it is also evident from Fig.~\ref{fig:rechi} the real part of the susceptibility does not have any significant weight around the small $\vQ$ regions which would explain why low energy FM spin fluctuations do not affect the dynamical susceptibility in Fig.~\ref{fig:imchi} for larger electron correlations.  

\begin{figure}
	\rput[tr](0.5\columnwidth,0.5\columnwidth){(a)}
	\includegraphics[width=0.49\columnwidth]{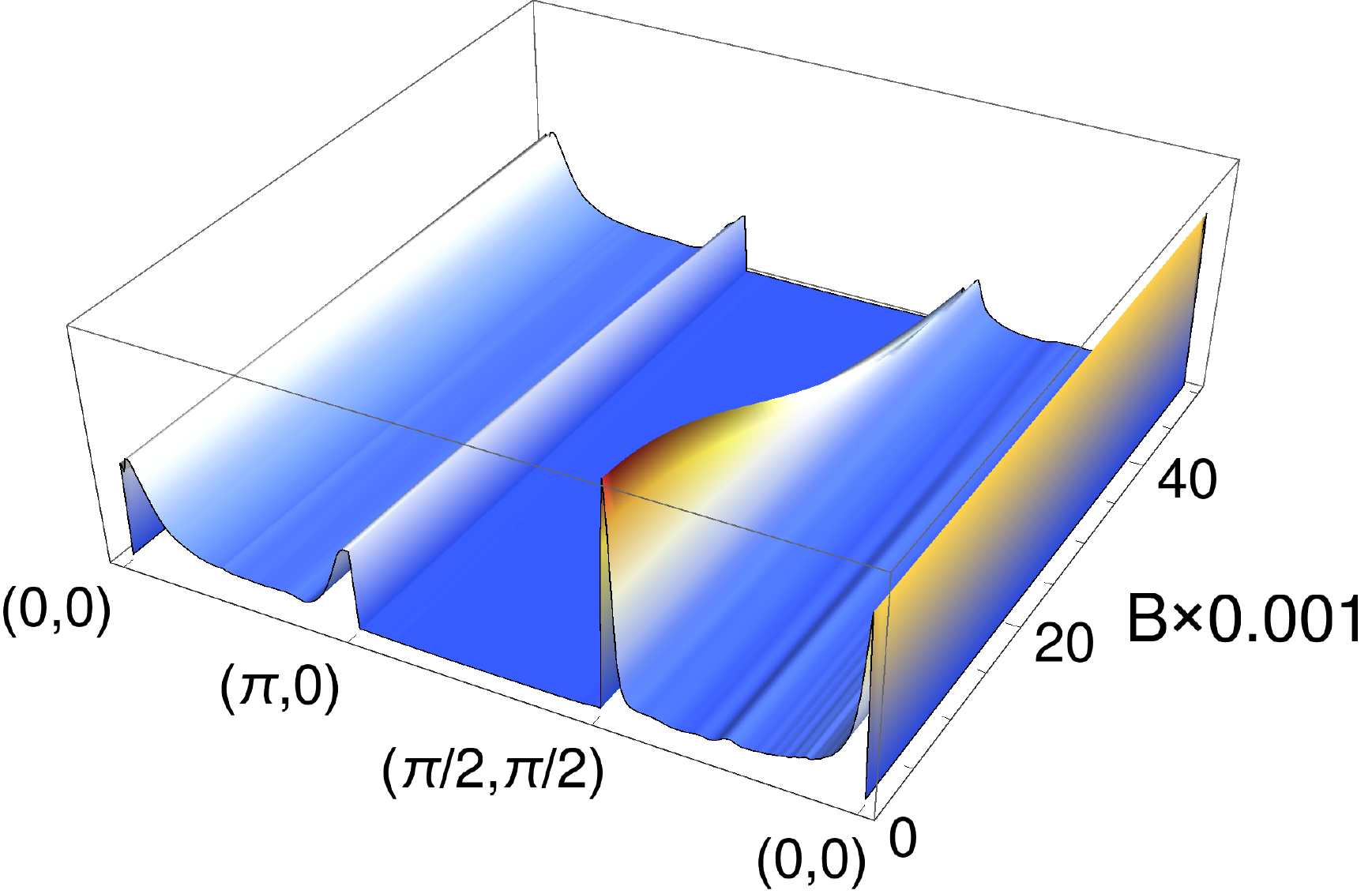}
	\rput[tr](0.5\columnwidth,0.5\columnwidth){(b)}
	\includegraphics[width=0.49\columnwidth]{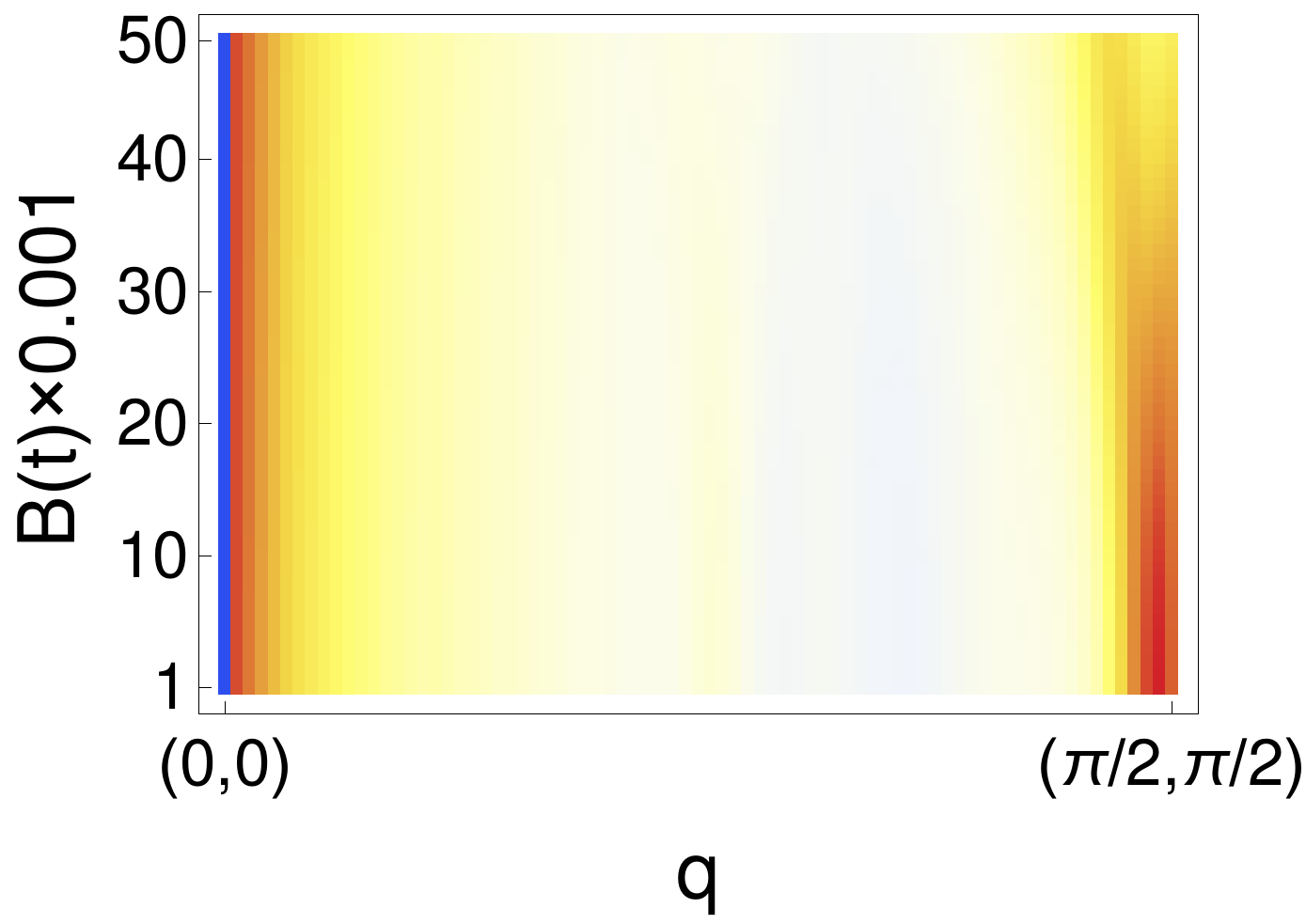}
	\caption{\label{fig:zee} Dynamical spin fluctuations in the presence of a Zeeman field in Sr327 for $x=1\%$ doping and $\omega=0.005$.(a) 3D plot showing the suppression of AFM fluctuation at $\vQ=(\pi/2,\pi/2)$. (b) 2D plot showing the development of incommensurability in the AFM spin fluctuations for larger magnetic fields.} 
\end{figure}

We now turn on the external Zeeman field along the $c$-direction to study its effect on the longitudinal spin fluctuations in the undoped compound. In Fig.~\ref{fig:zee}, we show the effect of an external magnetic field on the spin fluctuations for $x\sim 1\%$ and at $\omega=0.005$. From Fig.~\ref{fig:zee} two effects are immediately apparent. Firstly, we can see from Fig.~\ref{fig:zee} that an external magnetic field suppresses the spin fluctuations at the AFM wavevector and for a magnetic field of around $B\sim0.01$(in units of hopping t) the AFM and FM fluctuations become of comparable magnitude. Therefore for larger magnetic fields, the material properties would be influenced by the stronger competition between the AFM and FM fluctuations. Secondly, larger external magnetic fields cause the AFM fluctuations to become incommensurate from the $\vQ=(\pi/2,\pi/2)$ direction. The stronger competition between AFM and FM order in finite magnetic fields and development of incommensurate spin fluctuations have been observed in neutron scattering experiments\cite{ramos:2008,lester:2015}. These features are however not observed in doped Sr327 compound. For a doping of $x\sim 6\%$, the spin fluctuations are not significantly affected by a c-axis Zeeman field, the fluctuations do not become incommensurate like in the undoped compound, and unlike in the undoped compound the ferromagnetic fluctuations do not compete with the AFM fluctuations at finite magnetic field (see Supplementary Fig.~S3).  

In summary we find that with reduced octahedral rotation and increasing electron doping the spin fluctuation wave vector in Sr327 shifts from $(\pi/2,\pi/2)$ to the $(\pi,0)$ and simultaneously gets enhanced. The enhancement of the spin fluctuations is strong enough to lead to a $\vQ=(\pi,0)$ magnetic state as the system approaches the stoner instability with increasing doping. In the presence of a Zeeman field, we find that the magnetic field suppresses the dominant $(\pi/2,\pi/2)$ spin fluctuations in the undoped compound and lead to a stronger competition between the AFM and FM fluctuations for larger fields. Further the magnetic field also causes the spin fluctuations to become more incommensurate. These observations for the doped Sr327 and in the presence of a Zeeman field are in agreement with neutron scattering experiments. 

{\it Acknowledgement} --
We are grateful for valuable discussions with P. Aynajian and Michael Lawler.
This work has been supported by a start up fund from Binghamton University.

\newpage
\section{Supplementary Material}

\allowdisplaybreaks

\setcounter{equation}{0}
\renewcommand{\theequation}{S\arabic{equation}}
\setcounter{figure}{0}
\renewcommand{\thefigure}{S\arabic{figure}}

\maketitle
\section{Tight Binding Hamiltonian}
The tight binding Hamiltonian in the bilayer model contains a contribution from intra layer hoppings without octahedral rotation given by,
\be
\hat{A_s}(\vk)=
\left(\begin{array}{lll}
	\epsilon_{\vk}^{yz} & \epsilon_{\vk}^{off}+is\lambda & 0\\
	\epsilon_{\vk}^{off}-is\lambda & \epsilon_{\vk}^{xz} & 0\\
	0 & 0 & \epsilon_{\vk}^{xy}
\end{array}\right)
\ee
\begin{figure}
	\includegraphics[width=0.99\columnwidth]{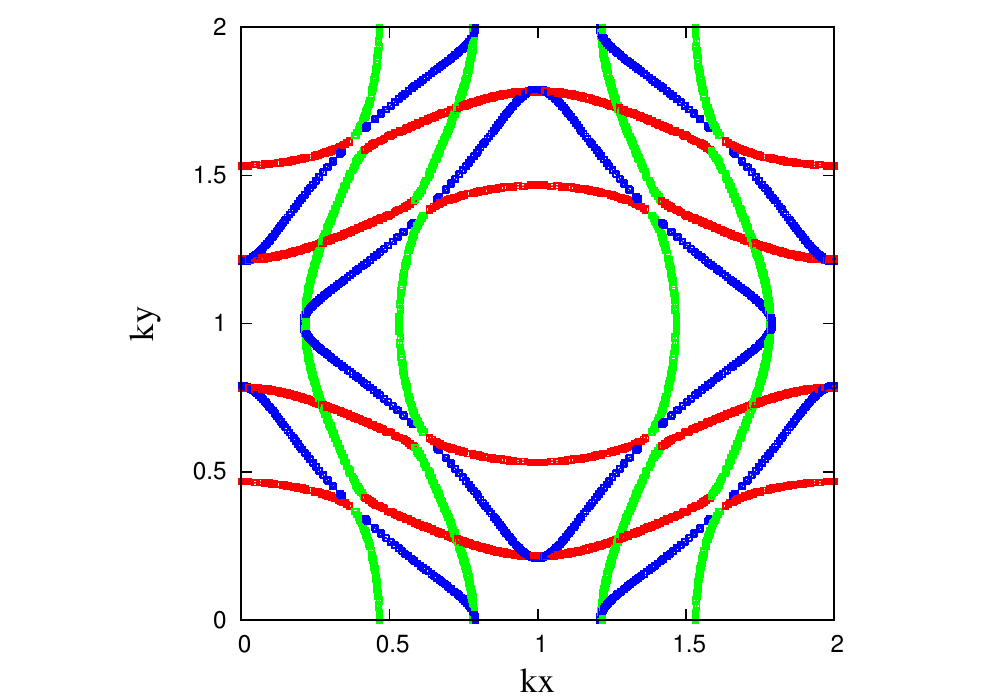}
	\caption{\label{fig:fs} The Fermi surface for $x\sim 1\%$ in the tetragonal unit cell. The colors represent the dominant orbital content with $d_{yz}$: red, $d_{xz}$: green, $d_{xy}$: blue. }
\end{figure}

where the matrix elements are given by,

\bea
&&\epsilon_{\vk}^{xz}=-2t_1\cos(k_x)-2t_2\cos(k_y),\\
&&\epsilon_{\vk}^{xy}=-2t_3(\cos(k_x)+\cos(k_y))-4t_4\cos(k_x)\cos(k_y)\nn\\
&&-2t_5(\cos(2k_x)+\cos(2k_y))\\
&&\epsilon_{\vk}^{off}=-4t_6\sin(k_x)\sin(k_y)
\eea

The hopping matrix elements are $t_1=t_3=0.5$, $t_2=0.05$, $t_4=0.1$, $t_5=-0.03$, and $t_6=0.05$. The spin orbit coupling term is given by $\lambda=0.1$. Note that all hopping are in units of hopping $t$. 

The interlayer and staggered hopping elements are also $3\times3$ matrices given by, 
\bea
&&\hat{G}(\vk)=
\left(\begin{array}{lll}
	0 & t_{rot}\gamma_{\vk} & 0\\
	-t_{rot}\gamma_{\vk} & 0 & 0\\
	0 & 0 & 0
\end{array}\right)\\
&&\hat{B_1}=
\left(\begin{array}{lll}
	-t_{\perp} & 0 & 0\\
	0 & -t_{\perp} & 0\\
	0 & 0 & 0
\end{array}\right)\\
&&\hat{B_2}=
\left(\begin{array}{lll}
	0 & t_{INT}^{\perp} & 0\\
	-t_{INT}^{\perp} & 0 & 0\\
	0 & 0 & 0
\end{array}\right)
\eea

In the above $\gamma_{\vk}=\cos(k_x)+\cos(k_y)$.
In Fig.~\ref{fig:fs} we have shown the Fermi surface for $x\sim 1\%$. Note that the dominant $(\pi/2,0)$ nesting in a tetragonal unit cell for the undoped material is primarily contributed by the $d_{xy}$ band (blue) whereas the dominant $(\pi/2,\pi/2)$ nesting for the doped compound are contributed by the $d_{xz}/d_{yz}$ bands (red/green).
\begin{figure}[b]
	\includegraphics[width=0.99\columnwidth]{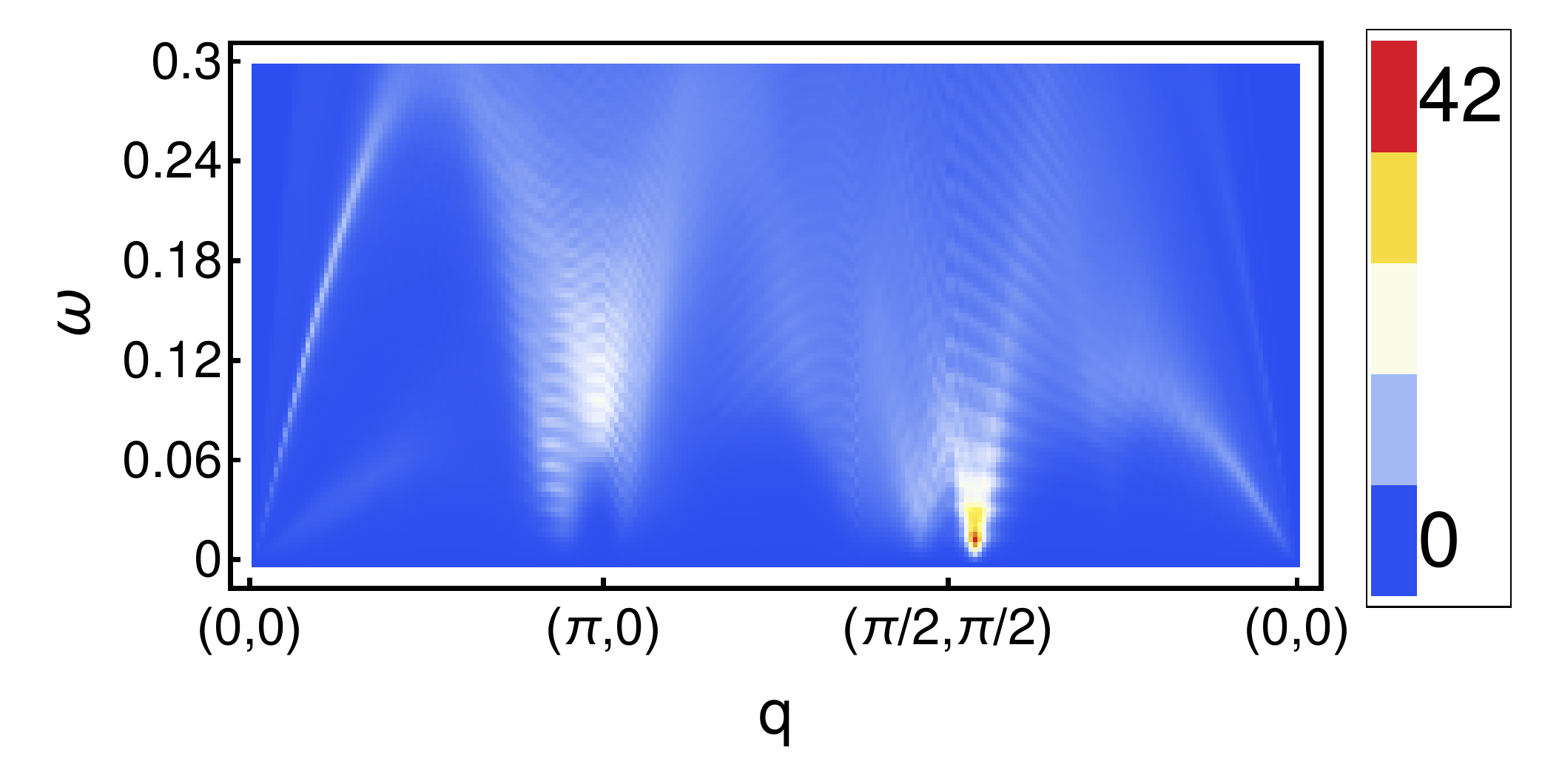}
	\caption{\label{fig:isoval} Imaginary part of spin susceptibility $\chi''_{RPA}(\vq,\omega)$ for $t_{rot}=0.36$ and chemical potential $\mu$ adjusted to give $x\sim 1\%$.}
\end{figure}

In Fig.~\ref{fig:isoval} we show the low energy spin fluctuations for a reduced hopping corresponding $t_{rot}=0.36$ but chemical potential adjusted to correspond to isovalent dopant effect thus giving a doping of $x\sim 1\%$. Note that this leads to incommensurability in the dominant spin fluctuations.
\begin{figure}
	\includegraphics[width=0.99\columnwidth]{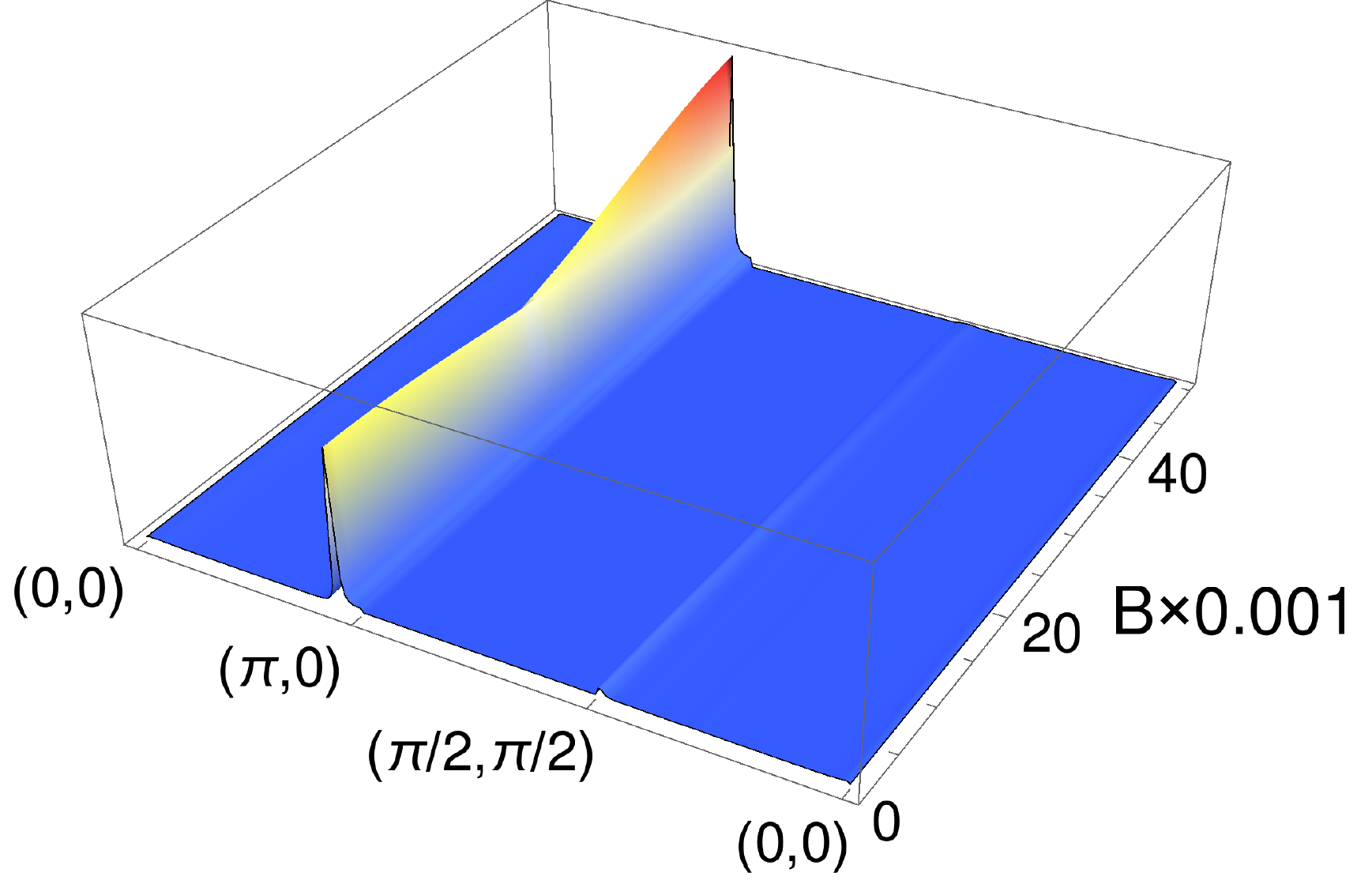}
	\caption{\label{fig:dopdepb} Dynamical spin fluctuations in the presence of a Zeeman field in Sr327 for $x=6\%$ doping and $\omega=0.005$.}
\end{figure}

In Fig.~\ref{fig:dopdepb} we show the magnetic field dependence of low energy spin fluctuations for a doping level of $x\sim 6\%$. We find that the peak does not change much until at much larger magnetic field where it is enhanced unlike the undoped compound.The susceptibility peak stays commensurate at the $(\pi,0)$ point, and finally
there are no competing ferromagnetic fluctuations, but this is simply to do with the fact that we are now closer to the stoner instability.

\end{document}